\documentstyle[11pt,newpasp,twoside,epsf]{article}
\def\edcomment#1{\iffalse\marginpar{\raggedright\sl#1\/}\else\relax\fi}
\reversemarginpar

\begin{document}

    \title{Some aspects of the chemical evolution of $\,^4{\rm He}$ in the
           Galaxy: the He/H radial gradient and the $\Delta Y/\Delta Z$
           enrichment ratio}
    \author{W. J. Maciel}
    \affil{IAG/USP - Av. Miguel Stefano 4200 - CEP 04301-904 S\~ao Paulo SP,
                     Brazil}

    \begin{abstract}
Two aspects of the chemical evolution of $\, ^4$He in the Galaxy 
are considered on the basis of a sample of disk planetary nebulae 
by the application of corrections due to the contamination of $\,^4$He 
from the progenitor stars. First, the He/H radial gradient is analyzed, 
and then, the helium to heavy element enrichment ratio is determined
for metallicities up to the solar value.

    \end{abstract}

    \section{Introduction}

The study of radial abundance gradients and of the enrichment ratio
between helium and the heavy elements ($\Delta Y/\Delta Z$) can be 
performed on the basis of photoionized nebulae, comprising HII  regions 
and planetary nebulae (PN, see for example the reviews by Peimbert 1990 
and Maciel 1997). The former have the advantage of representing the 
interstellar composition directly, while for the latter the influence 
of the contamination from the  progenitor stars should be taken into 
account, particularly for  those elements that are dredged up to the 
outer layers of the progenitor stars, such as helium and nitrogen.

In the case of helium, the contamination from the evolution of the
progenitor stars has been recently determined on the basis of
different theoretical models (cf. van den Hoek \& Groenewegen 1997,
Boothroyd \& Sackmann 1999), so that it is now possible to apply
individual corrections to the observed PN abundances in order to
obtain improved values of the He/H radial gradient and of the 
$\Delta Y/\Delta Z$ ratio. In the present work, these recent 
calculations are used in order to estimate the He contamination 
for a sample of disk planetary nebulae recently studied by Maciel
\& Quireza (1999). The application of  corrections to the observed 
abundances leads to a new determination of the He/H radial gradient
$d$(He/H)/$dR$ and of the slope $\Delta Y/\Delta Z$. 

    \section{The He/H radial gradient and the $\Delta Y/\Delta Z$ ratio}

    \subsection{The He/H radial gradient}

The O/H radial gradient is now well known, and can be derived for
HII regions, planetary nebulae, hot stars, etc. It amounts roughly
to $d\log ({\rm O/H})/dR \simeq -0.07$ dex/kpc, and similar values
have been obtained for other elements, such as S and Ar (Maciel 1997, 
Maciel \& Quireza 1999). The existence of a He/H gradient is much 
more uncertain, and results obtained both from HII regions and planetary
nebulae show a very large uncertainty, as seen in Table 1 (see
Esteban \& Peimbert 1995 for a review).

    \begin{table}
    \caption{Determinations of the He/H radial gradient}
    \begin{tabular}{llll}
    \tableline
Reference & PN/HII & $d$(He/H)/$dR$ & $d\log$(He/H)/$dR$  \\   
     &        & (kpc$^{-1}$)   & (dex/kpc) \\   
    \tableline
D'Odorico et al. (1976)          & PN  & $-0.007$  & $-0.03$  \\
Peimbert \& Serrano (1980)       & PN  & $-0.005$  & $-0.02$  \\
Shaver et al. (1983)             & HII & $-0.0003$ & $-0.001$ \\
Fa\'undez-Abans \& Maciel (1986) & PN  & $-0.005$  & $-0.02$  \\
Pasquali \& Perinotto (1993)     & PN  & $-0.003$  & $-0.009$ \\
Amnuel (1993) [min]              & PN  & $-0.0005$ & $-0.002$ \\
Amnuel (1993) [max]              & PN  & $-0.006$  & $-0.026$ \\
Maciel \& Chiappini (1994)       & PN  & $-0.0004$ & $-0.002$ \\
Esteban et al. (1999)            & HII & $-0.001$  & $-0.004$ \\
    \tableline
    \tableline
    \end{tabular}
    \end{table}

All determinations involving PN until now have {\it not} taken into 
account the He contamination by the progenitor stars, so that
it is interesting to revise the values of the He/H gradient by
considering the amount of He produced and dredged up to the
outer layers of the stars and eventually shed into the nebulae.

\subsection{The $\Delta Y/\Delta Z$ enrichment ratio}

The helium to metals enrichment ratio $\Delta Y/\Delta Z$ is generally 
determined adopting a linear variation for the helium abundance by mass 
with the metallicity $Z$ of the form $Y(Z) = Y_p + (\Delta Y/\Delta Z) 
\, Z $, where $Y_p$ is the pregalactic helium abundance, as proposed by 
Peimbert and Torres-Peimbert (1974, 1976). Photoionized nebulae such as 
HII regions and blue compact  galaxies are generally used (cf. Izotov et 
al. 1997, Thuan \& Izotov 1998, Esteban et al. 1999), and recent work has 
also taken into account the fine structure in the main sequence of nearby 
stars (Pagel \& Portinari 1998). Results are generally in the range 
$2 \leq \Delta Y/\Delta Z \leq 6$, as shown in Table~2.

    \begin{table}
    \caption{Determinations of the $\Delta Y/\Delta Z$ ratio}
    \begin{tabular}{lll}
    \tableline
Reference & object  & $\Delta Y/\Delta Z$  \\   
    \tableline
D'Odorico et al. (1976)    & PN       & $2.95$        \\
Peimbert \& Serrano (1980) & PN       & $2.2-3.6$     \\
Maciel (1988)              & PN       & $3.5\pm 0.3$  \\
Pagel et al. (1992)        & HII      & $6.1\pm 2.1$  \\
Chiappini \& Maciel (1994) & PN/HII   & $3.4-5.6$     \\
Pagel \& Portinari (1998)  & MS stars & $3\pm2$       \\
Thuan \& Izotov (1998)     & HII      & $2.3\pm 1.0$  \\
Esteban et al. (1999)      & HII      & $1.9-3.9$     \\
    \tableline
    \tableline
    \end{tabular}
    \end{table}

\noindent
Maciel (1988) determined $Y_p$ and $\Delta Y/\Delta Z$ using a sample 
of type~II PN (Peimbert 1978), and assuming that the He contamination 
from the central star was negligible. Chiappini \& Maciel (1994)
made a first attempt at including this contamination in a systematic
way, and added a term $\Delta Y_s$ to the $Y(Z)$ relation, which 
can be written as

    \begin{equation}
Y = Y_p + {\Delta Y\over \Delta Z} \ Z + \Delta Y_s \ .
    \end{equation}

\noindent
The stellar contribution $\Delta Y_s$ was obtained on the basis 
of some calculations by Boothroyd (private communication). The adopted 
values were $\Delta Y_s = 0.0$, 0.008, 0.015 and 0.022, which 
were applied to all nebulae in the sample, so that any differences 
in their {\it individual} behaviour were lost. 

\section{He contamination from the PN progenitor stars}

The He excess by mass $\Delta Y_s$ can be estimated from the yields
of the intermediate mass stars as a function of the stellar mass 
on the main sequence $M_{MS}$ and total metallicity $Z$, adopting
$Z = 0.020$, which is appropriate for type II PN. The stellar mass
on the main sequence can be obtained by an initial mass-final mass
relation as a function of the PN core mass $M_c$, which can in
principle be determined from the observed nebular abundances,
particularly the N/O ratio.

In this work, we have adopted the recent yields by van den Hoek 
\& Groenewegen (1997), using as comparison the calculations by
Boothroyd \& Sackmann (1999). The former includes all three dredge 
up processes that occur in the late stages of the intermediate 
mass stars, and generally produce larger yields than the latter,
as can be seen in Figure~1. We have assumed that most of the 
He excess that are dredged up to the surface is mixed up in the 
outer layers and ejected into the nebulae, so that our derived 
$\Delta Y_s$ is the largest possible correction for a given mass.

    \begin{figure}
    \plotfiddle{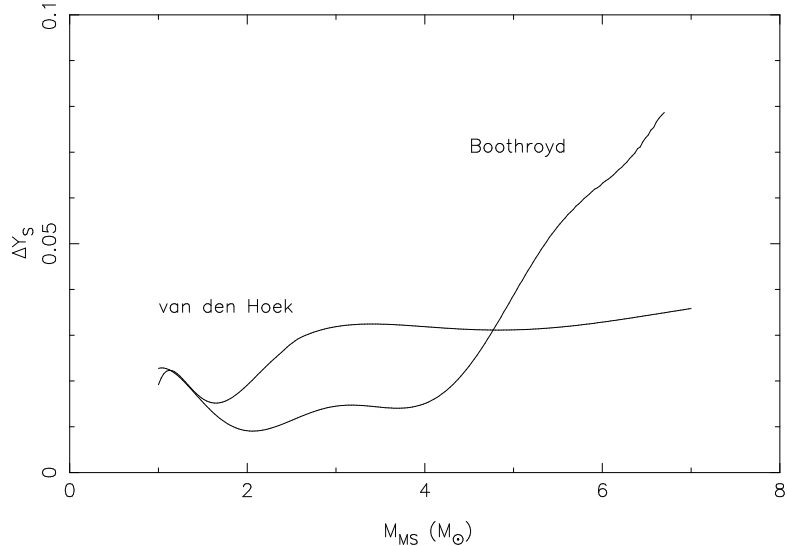}{7 cm}{0}{60}{60}{-170}{-245}
    \caption{He contamination from PN progenitor stars.}
    \end{figure}

In view of the uncertainties on the PN central star masses, we have
adopted two different calibrations, referred to as {\it high mass}
and {\it low mass} calibrations, respectively. For the {\it high mass
calibration}, we adopted the $M_c$ $\times$ N/O relation recently
proposed by Cazetta \& Maciel (2000),

    \begin{equation}
    M_c = a + b \ \log({\rm N/O}) + c \ [\log({\rm N/O})]^2 
    \end{equation}

\noindent
where $a = 0.689$, $b = 0.056$ and $c = 0.036$ for $-1.2 \leq \log 
({\rm N/O}) \leq -0.26$ and $a = 0.825$, $b = 0.936$ and $c = 1.439$ 
for $\log ({\rm N/O}) > -0.26$.  The average initial mass-final mass 
relation was taken from the gravity distance work of Maciel \& Cazetta 
(1997), and can be written as

    \begin{equation}
    M_c = a_0 + a_1 \ M_{MS} + a_2 \ M_{MS}^2 + a_3 \ M_{MS}^3
    + a_4 \ M_{MS}^4
    \end{equation}

\noindent
where $a_0 = 0.5426$, $a_1 = 0.02093$,  $a_2 = -0.01122$,
$a_3 = 0.00447$ and  $a_4 = -0.0003119$. This calibration leads to
core masses $M_c \geq 0.67 M_\odot$ and main sequence masses 
$M_{MS} \geq 3\, M_\odot$, showing a good agreement with the
results from NLTE model atmospheres of M\'endez et al. (1988, 1992).

For the {\it low-mass calibration} we can still use equation~2, replacing
the coefficients by $a = 0.7242$, $b = 0.1742$ and $c = 0$ for 
$\log ({\rm N/O}) \leq -0.26$. For the initial mass--final mass
relation we have the coefficients $a_0 = 0.4877$, $a_1 = 0.0623$,
$a_2 = a_3 = a_4 = 0$. This calibration produces masses in the
ranges $M_c \geq 0.55\, M_\odot$ and $M_{MS} \geq 1 M_\odot$,
which agree with the recent determinations of PN central star
masses of Stasi\'nska et al. (1997) and with the masses of type II
PN originally proposed by Peimbert (1978). Both $M_c \times $ N/O
calibrations are shown in figure~2, and the initial mass-final
mass relation for the high-mass calibration can be seen in the
figure~1 of Maciel \& Cazetta (1997).

    \begin{figure}
    \plotfiddle{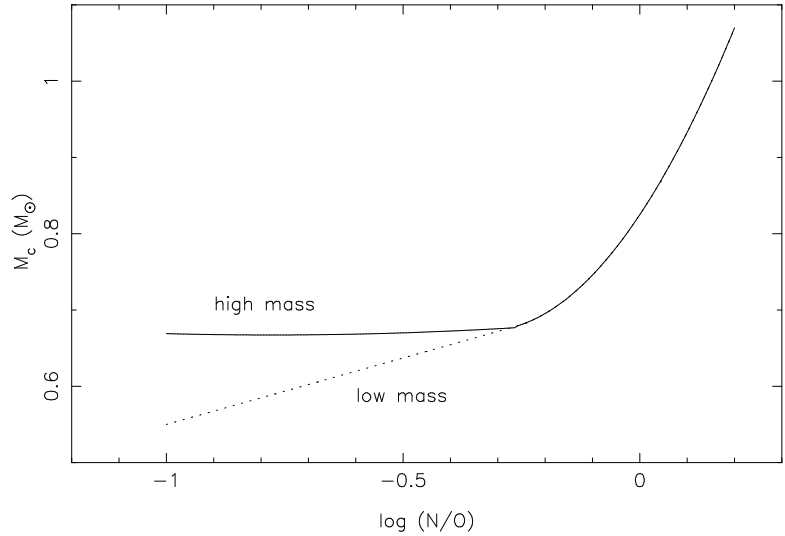}{7 cm}{0}{60}{60}{-170}{-245}
    \caption{The adopted $M_c \times $ N/O calibrations.}
    \end{figure}

\section{Results and discussion}

We have applied the corrections outlined in the previous section
to a sample of disk planetary nebulae in order to derive the
He/H radial gradient and the $\Delta Y/\Delta Z$ enrichment ratio.
For details on the objects and abundances, the reader is referred to
Maciel \& Quireza (1999) and Maciel (2000).

\subsection{The He/H radial gradient}

The average He/H abundances by number of atoms are shown in Table~3,
using the van den Hoek \& Groenewegen (1997) and Boothroyd \& Sackmann
(1999) data, both for the high- and low-mass calibrations.

    \begin{table}
    \caption{The He/H average abundances}
    \begin{tabular}{ll}
    \tableline
                                    &  He/H \\   
    \tableline
Uncorrected abundances              & $0.106 \pm 0.003$ \\
Boothroyd, low mass calibration     & $0.100 \pm 0.003$ \\
Boothroyd, high mass calibration    & $0.100 \pm 0.003$ \\
van den Hoek, low mass calibration  & $0.094 \pm 0.003$ \\
van den Hoek, high mass calibration & $0.091 \pm 0.003$ \\
    \tableline
    \tableline
    \end{tabular}
    \end{table}

\noindent
The derived gradients are negligible, and can be written as
$d{\rm (He/H)}/dR = 0.0000 \pm 0.0004$, irrespective of the calibration
used, as shown for example in Figure~3 for the van den Hoek \& 
Groenewegen (1997) data with the low mass calibration. We 
have adopted $R_0 = 7.6$ kpc as in Maciel \& Quireza (1999). 

    \begin{figure}
    \plotfiddle{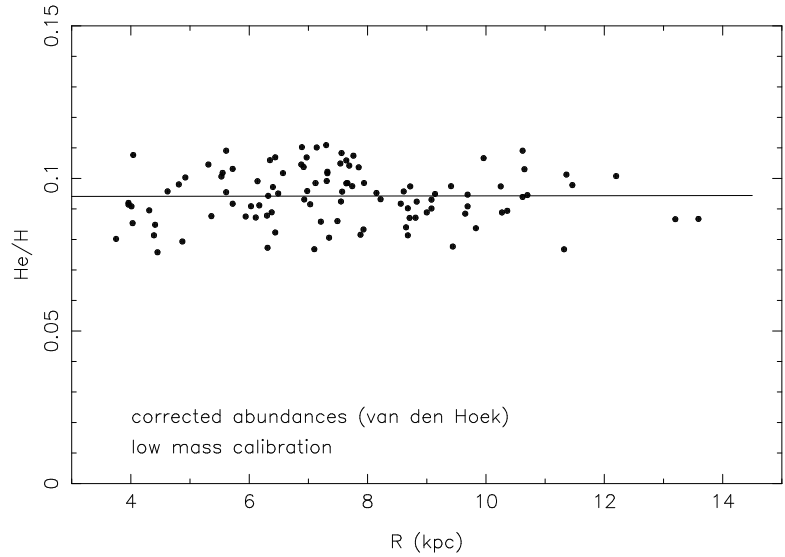}{7 cm}{0}{60}{60}{-170}{-245}
    \caption{He/H abundances as a function of the galactocentric 
             distance for the van den Hoek and Groenewegen (1997)
             data with the low mass calibration.}
    \end{figure}

It can be seen that the correction procedure simply reduces the average 
He/H abundances, so that stars with different masses are probably 
scattered homogeneously in the whole range of galactocentric 
distances, thus destroying any systematic variations. In view
of the uncertainties involved in the abundances (and distances),
it is unlikely that any He/H radial gradient could be presently  
detected from planetary nebulae, and previous determinations 
were probably affected by the use of small samples.

On the other hand, an upper limit to the He/H gradient can be derived
on the basis of the total dispersion $\sigma_d$ observed. Taking  
$\sigma_d \simeq 0.04$, we have $\vert d({\rm He/H})/dR\vert  < 
\sigma_d/ \Delta R \simeq  0.004 \ {\rm kpc}^{-1}$, or 
$d\log({\rm He/H})/dR \simeq -0.02$ dex/kpc. Therefore, 
any existing He/H radial gradient should be lower than the O/H 
gradient by at least a factor of 3. This conclusion is in agreement 
with the small gradients derived for galactic HII regions by Esteban 
et al. (1999) and with some recent chemical evolution models 
for the Galaxy (Chiappini et al. 1997, Chiappini, this conference).

\subsection{The $\Delta Y/\Delta Z$ enrichment ratio}

In order to apply the correction procedure to the PN sample, we
have considered the O/H abundances as representative of the total 
metallicity, adopting $Z \simeq 25 \ $O/H (Peimbert 1990, Chiappini \& 
Maciel 1994). Since the pregalactic abundance can be better 
determined on the basis of very low metallicity objects, we 
have taken $Y_p$ as a parameter, with the values $Y_p = 0.23$ 
and $Y_p = 0.24$ (see for example Olive et al. 1999, Izotov et
al. 1999, Steigman, this conference). 

We have taken into account all PN in our sample having metal 
abundances up to $10^6\, {\rm O/H} \simeq 700$, which corresponds 
approximately to the solar value, $\epsilon({\rm O}) = \log {\rm (O/H)} 
+ 12 = 8.83$ (Grevesse \& Sauval 1998), or $Z \simeq 0.017$. As pointed 
out by Chiappini \& Maciel (1994), He abundances of PN show some tendency 
to flatten out for very large $Z$, where a more sophisticated relation
than eq. (1) would be needed. Moreover, some large metallicity 
PN have larger than average N/O ratios, so that the corrections to the 
He abundance are also larger and more uncertain. 

The main results are shown in Table 4 and in figures~4 and 5, where
the PN are shown as filled circles. Figure~4 shows the uncorrected 
abundances and fits, and figure~5 shows the results using the 
corrections according to the van den Hoek and Groenewegen data both
for the high- and low-mass calibrations. In both figures, the straight 
lines are least squares fits using $Y_p = 0.23$ (dashed lines)  
and $Y_p = 0.24$ (solid lines).

    \begin{table}
    \caption{The $\Delta Y/\Delta Z$ ratio}
    \begin{tabular}{ll}
    \tableline
                & $\Delta Y/\Delta Z$  \\   
    \tableline
$Y_p = 0.23$           &                 \\
uncorrected            & $5.60 \pm 0.22$ \\
Boothroyd low mass     & $4.42 \pm 0.20$ \\
Boothroyd high mass    & $3.95 \pm 0.19$ \\
van den Hoek low mass  & $3.59 \pm 0.19$ \\
van den Hoek high mass & $2.87 \pm 0.17$ \\
                       &                 \\
$Y_p = 0.24$           &                 \\
uncorrected            & $4.73 \pm 0.20$ \\
Boothroyd low mass     & $3.55 \pm 0.19$ \\
Boothroyd high mass    & $3.08 \pm 0.17$ \\
van den Hoek low mass  & $2.73 \pm 0.17$ \\
van den Hoek high mass & $2.01 \pm 0.16$ \\
    \tableline
    \tableline
    \end{tabular}
    \end{table}

    \begin{figure}
    \plotfiddle{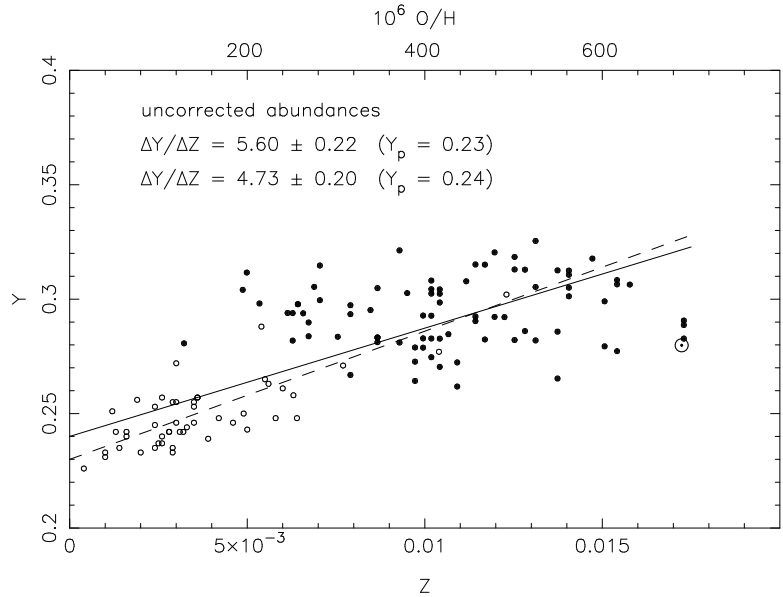}{7 cm}{0}{60}{60}{-170}{-245}
    \caption{Uncorrected He abundances by mass $Y$ for PN (solid 
             circles) as a function of O/H (top axis) and $Z$ 
             (bottom axis). Also shown are the Sun ($\odot$) and HII 
             regions (empty circles). The straight lines are least 
             squares fits for $Y_p = 0.23$ (dashed line) and $Y_p = 0.24$ 
             (solid line).}
    \end{figure}

For comparison purposes, figures~4 and 5 also include the Sun ($\odot$) 
and the HII regions and metal poor blue compact galaxies 
(empty circles) from the compilation of Chiappini \& Maciel (1994).
These objects have {\it not} been taken into account in the determination 
of the linear fits, and are included for comparison only.

    \begin{figure}
    \plotfiddle{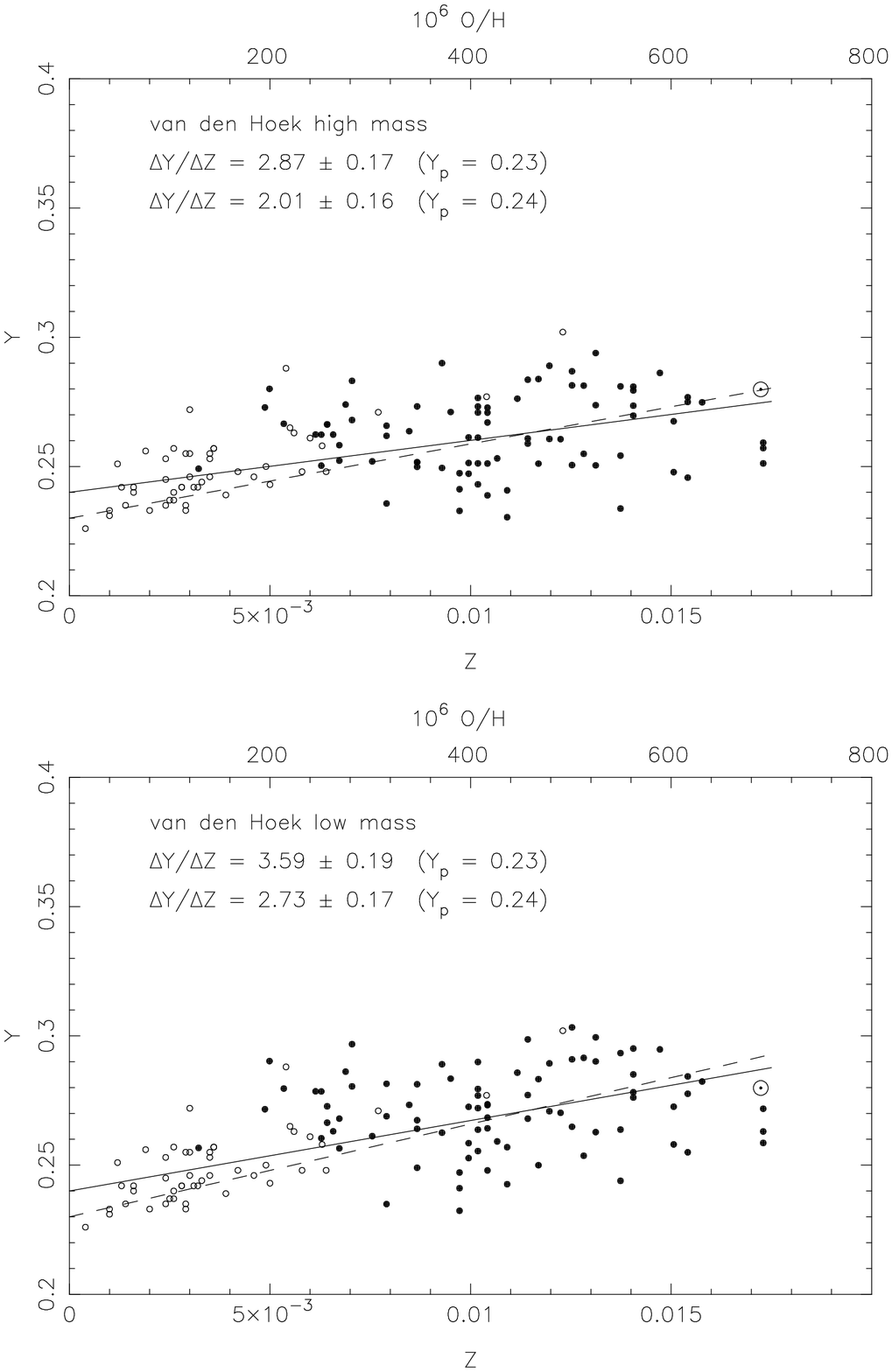}{16 cm}{0}{60}{60}{-170}{0}
    \caption{The same as figure~4 using corrected abundances 
            according to the van den Hoek \& Groenewegen (1997)
            data for the high mass calibration (top panel) and low mass 
            calibration (bottom panel).}
     \end{figure}

It can be seen that the scatter in the PN data is somewhat higher,
which is partially due to the higher uncertainties in the abundances 
and also to the adopted correction procedure, so that future improvements
along these lines would be desirable. The higher uncertainties in the 
PN abundances and the lower He/H gradient are probably responsible 
for the lack of a He/H $\times$ $R$ correlation, while some  
correlation between $Y$ and $Z$ (or O/H) can be observed.

The application of the correction procedure reduces the average 
He abundances and the $\Delta Y/\Delta Z$ ratio, and also 
decreases the uncertainty of the derived slopes. The $\Delta Y/\Delta Z$ 
ratio decreases from the range 4.7 -- 5.6 to values in the range 
2.8 -- 3.6 for $Y_p = 0.23$ and 2.0 -- 2.8 for $Y_p = 0.24$. These  
results are closer to recent independently derived ratios, as seen 
in Table~2 (see also Thuan, this conference), and to the predictions 
of theoretical models (Allen et al. 1998, Chiappini et al. 1997), 
particularly for the high mass calibration using the van den Hoek 
\& Groenewegen (1997) data. 

The average uncertainties in the derived slopes are smaller as 
a consequence of the fact that a large number of objects has been 
included, with a larger metallicity spread.
An average including  both calibrations would give $\Delta Y/\Delta Z = 
3.2 \pm 0.5$ for $Y_p = 0.23$ and  $\Delta Y/\Delta Z = 2.4 \pm 0.5$ 
for $Y_p = 0.24$, which can be compared with the uncertainties shown
in Table~2. 

Finally, the apparent continuity between the low metallicity
and high metallicity objects of figure~5 suggests that their
chemical evolution may not have been very different. 
In fact, the chemical evolution of a system is basically 
defined by its initial mass function (IMF) and star formation history 
(SFH). The IMF is now believed to be universal (Maciel \& Rocha-Pinto
1998, Padoan et al. 1997). Blue compact galaxies have bursts of star
formation, a feature that has been recently reinforced in
our own Galaxy, as shown by Rocha-Pinto et al. (2000) on the
basis of chromospheric ages. Therefore, the similarity in the chemical 
evolution of the different systems shown in figure~5 is probably 
not surprising.

\bigskip\noindent
   {\it Acknowledgements.} This work was partially supported by
   CNPq and FAPESP.

\end{document}